\documentclass[english,aps,reprint,showpacs,titlepage,longbibliography]{revtex4-2}

\usepackage[T1]{fontenc}	% should generally be included for better accented-word behavior
\usepackage[latin9]{inputenc}	% should generally be included for better accent behavior
\usepackage{geometry}		% for controlling page margins
\usepackage{graphicx}
\usepackage[above,below]{placeins}	% allows use of \FloatBarrier command to force section barriers
\usepackage{times}

\usepackage{hyperref}
\hypersetup{colorlinks=true,urlcolor=blue,citecolor=blue,linkcolor=blue}
\usepackage{enumitem}          % package for handling list formatting
\setlist{nosep}                 % Tightest spacing for lists. `noitemsep` is more relaxed

\usepackage{amsmath}
\usepackage{ragged2e}
\usepackage[english]{babel}
\usepackage{longfbox}
\usepackage{nicefrac}
\usepackage[capitalise]{cleveref}
\usepackage{braket}
\usepackage{microtype}

\usepackage{soul,xcolor}
\colorlet{BLACK}{black} % I hate LaTeX so much.
\sethlcolor{black}

\newcommand{\censor}[1]{#1}
\newcommand{\censorhref}[2]{\href{#1}{#2}}
\newcommand{\censorcite}[1]{\cite{#1}}
\newcommand{\censorace}[1]{#1}

\begin{document}

% \StopCensoring

\begin{titlepage}

  \title{
    Effectiveness of an online homework tutorial about changing basis in quantum mechanics
  }

  \author{Giaco Corsiglia}
  \author{Steven Pollock}
  \author{Bethany R. Wilcox}
  \affiliation{Department of Physics, University of Colorado, Boulder, Boulder, Colorado, 80309, USA}

  % \keywords{}

  \begin{abstract}
    Changing basis is a common task when solving quantum mechanical problems.  As part of a research project investigating student understanding of basis and change of basis in quantum mechanics, we developed a tutorial to support students in learning about basis in the context of spin-\nicefrac{1}{2} systems.  We have since created an interactive online version of the basis tutorial as part of a freely available suite of online quantum tutorials called ACE Physics (\href{https://acephysics.net}{acephysics.net}).  The ACE Physics tutorials include dynamic guidance elements and, unlike other tutorials, are intended for use outside the classroom without instructor facilitation.  After extensive study in an instructor-supported environment, we assigned the ACE Physics basis tutorial as homework in two semesters of upper-division quantum mechanics, and we report on the effectiveness of the activity based on pre-/post-testing and comparison of student exam performance with a similar semester that did not include the activity.  We find that the tutorial produces sufficient learning gains to justify continued assignment as a homework problem in our classes.
    \clearpage
  \end{abstract}

  \maketitle
\end{titlepage}

% Times New Roman 10 point font
\section{Introduction}

Quantum mechanical problems often require converting between bases.  Our collaboration's prior research explored student ideas about basis and change of basis and identified conceptual and procedural challenges that students encounter when learning these topics in a spins-first quantum mechanics course~\censorcite{BasisPaper1}.  We developed an on-paper tutorial---the Quantum Basis Tutorial (QBT)---to support student learning about basis~\footnote{The on-paper Quantum Basis Tutorial is available at~\censorcite{ACEQM}.}.  Our previous work described the development process for the QBT and presented preliminary evidence of its effectiveness~\censorcite{BasisPaper2}.

The QBT discusses basis in the context of spin-\nicefrac{1}{2} quantum systems.  In this context, changing basis refers, for example, to the calculation required to derive \cref{eq:psi-in-x} from \cref{eq:psi-in-z},
\begin{align}
  \ket{\psi}&=\frac{2}{\sqrt{5}}\ket{+} +\frac{1}{\sqrt{5}}\ket{-} \label{eq:psi-in-z}\\
  \ket{\psi}&=\frac{3}{\sqrt{10}}\ket{+}_x +\frac{1}{\sqrt{10}}\ket{-}_x  \label{eq:psi-in-x}
\end{align}
One method for performing this calculation involves projection inner products.  For example, the coefficient in front of $\ket{+}_x$ is equal to the inner product ${}_x{\braket{+|\psi}}$.  Other methods also work, but projection typically requires fewer algebraic steps and generalizes to higher dimensional Hilbert spaces.

As presented in \censorcite{BasisPaper2}, the QBT has two primary learning goals.  After completing the QBT, students should be able to:
\begin{enumerate}
  \item[LG1.] Recognize that changing basis does not change the state or the probabilities of any measurement on a state;
  \item[LG2.] Use projection as a method to change basis with and without prompting.
\end{enumerate}
The tutorial uses an analogy to two-dimensional Cartesian space.  It asks students to represent a Cartesian vector using Dirac notation, matrix notation, and graphically all in two different bases (i.e., two sets of rotated Cartesian axes) and to compare these representations.

We have since developed an interactive online version of the QBT as part of a growing suite of online quantum mechanics tutorials called \censorace{ACE Physics} (\censor{Adaptable Curricular Exercises for Physics})~\censorcite{ACEPhysics}.  Unlike the on-paper version of the QBT, the \censorace{ACE Physics} version is designed to be completed outside of class and without instructor facilitation.  We have assigned the \censorace{ACE Physics} QBT~\censorcite{ACEPhysicsQBT} on homework assignments in two semesters of upper-division quantum mechanics.  This paper presents a preliminary evaluation of the effectiveness of the QBT assigned in the novel \censorace{ACE Physics} format.

We consider the classroom to be the ideal tutorial environment, and we encourage faculty to consider running in-class tutorials for quantum mechanics using the variety of available resources (e.g., ~\cite{OSUQuantum,Kohlne2017Characterizing,QUILTS,PhET,UWQuantum,Emigh2020Research,ACEQM}), including possibly \censorace{ACE Physics} itself (the guidance elements may be useful in classes with high student-instructor ratios).  However, we recognize that in-class tutorials are not always an option, and we hypothesize that students will benefit more from an \censorace{ACE Physics} tutorial than from no tutorial at all.  This paper presents a preliminary test of that hypothesis for the QBT.

\section{\texorpdfstring{The \censorace{ACE Physics} Project}{The ACE Physics Project}}

Quantum mechanics is an essential course for undergraduate physics majors,  but the material is known to be conceptually challenging for students~\cite{Singh2015Review,Gire2015Structural,Passante2019Enhancing, Passante2020Investigating, Wan2017Student}.  Research-based \textit{tutorials}---guided-inquiry worksheets that assist students in constructing physics knowledge for themselves---have repeatedly proven to be a highly effective complement to lectures in physics courses, and are especially targeted at supporting students' conceptual understanding~\cite{Shaffer1992Research, Finkelstein2005Replicating, Keebaugh2019Improving, Singh2008Interactive, UWQuantum}.  Our collaboration has developed a collection of tutorials for upper-division quantum mechanics (QM) that has been adopted by numerous QM instructors nationwide~\censorcite{ACEQM}.  We call these the ``\censor{ACEQM} tutorials''.

Designed to be completed in groups with facilitators present to guide students, tutorials require institutional and instructional resources and buy-in: an effective implementation relies on trained facilitators~\cite{TutorialsInIntroductoryPhysicsGuide, Finkelstein2005Replicating}.  These requirements compete with trends of increased student-instructor ratios and use of online materials~\cite{NAP18312}.  In upper-division physics courses such as QM, tutorials often face additional barriers:  limited class time, lack of a recitation section, and greater instructor autonomy, which makes it difficult to establish a tradition of running tutorials every semester.  The \censorace{ACE Physics} project aims to side-step these barriers by adapting tutorials to an online format that may be easily administered outside of class and does not require real-time  instructor intervention or a dedicated physical space.

The freely available \censorace{ACE Physics} website (\censorhref{https://acephysics.net}{acephysics.net}) hosts interactive online versions of all of the existing \censor{ACEQM} tutorials.  Whereas the original \censor{ACEQM} tutorials were designed to be done by students working on paper and---more significantly---in groups, in the classroom, and with instructors present, the \censorace{ACE Physics} versions are designed to be done by students working on their computers, outside of class, without instructors present, and (possibly) alone.  These differences require the \censorace{ACE Physics} tutorials to vary meaningfully from their on-paper counterparts.

We intend for our tutorials---both on-paper and on \censorace{ACE Physics}---to provide formative (as opposed to summative) experiences for students.  We do not grade tutorials (except perhaps for participation credit), and aim to provide space for students to comfortably engage with and refine their ideas regardless of initial correctness.  A typical on-paper tutorial presents a sequence of short, mostly conceptual questions about a given topic.  A typical \censorace{ACE Physics} tutorial does the same, but intersperses guidance elements based on student responses.  Although we frequently ask open-ended questions, we do \textit{not} attempt to parse students' written responses; dynamic guidance logic is based on multiple-choice and numerical responses only.   Example guidance elements include: hints, follow-up questions, messages prompting students to compare two of their answers that logically disagree, or messages explicitly telling students if they are correct or incorrect.  To some extent, we design \censorace{ACE Physics} guidance elements to model the conversations we have with students when facilitating in-class tutorials. 

\section{Methods}

This study includes data from three semesters---Fall 2018, Fall 2021, and Spring 2022---of the upper-division quantum mechanics course at a large, R1 university.  All three semesters were taught by PER faculty using similar curricular materials~\censorcite{ACEQM} and the text by McIntyre~\cite{McIntyre2012}, which adopts a \textit{spins-first} approach to introductory quantum mechanics.  A spins-first course presents the postulates of quantum mechanics in the context of spin-\nicefrac{1}{2} systems before discussing position-space wave functions in the second half of the semester.

All three semesters included three hours a week of in-person lecture (which included clicker questions), a weekly homework problem set, and three exams.  The two fall semesters were taught by author \censor{SP} and both included an optional, weekly, in-person recitation section during which on-paper \censor{ACEQM} tutorials were run.  The tutorial recitations were attended by roughly 30\% of the class in each semester.  The on-paper QBT was never administered during these recitations.  Other than the QBT, author \censor{SP} used nearly identical instructional materials in both fall semesters.   The spring semester, taught by author \censor{BW}, did not include a recitation section, but on-paper tutorials (excluding the QBT) were occasionally facilitated during lecture periods.

\textbf{Administration of the \censorace{ACE Physics} QBT:}
The \censorace{ACE Physics} QBT~\censorcite{ACEPhysicsQBT} was assigned as a homework problem on the fourth weekly problem set
in both the Fall 2021 and Spring 2022 semesters.  The QBT was not administered in any form during the Fall 2018 semester.  The version of the \censorace{ACE Physics} QBT used in these semesters consists of only three of the five pages of the full QBT described in~\censorcite{BasisPaper2}.  In both semesters, students' tutorial submissions were not graded for correctness.  Collaboration on homework assignments---including \censorace{ACE Physics}---was encouraged in both semesters.

\begin{figure}
  {%
  \setlength{\fboxsep}{0pt}%
  \fbox{\includegraphics[width=.99\linewidth]{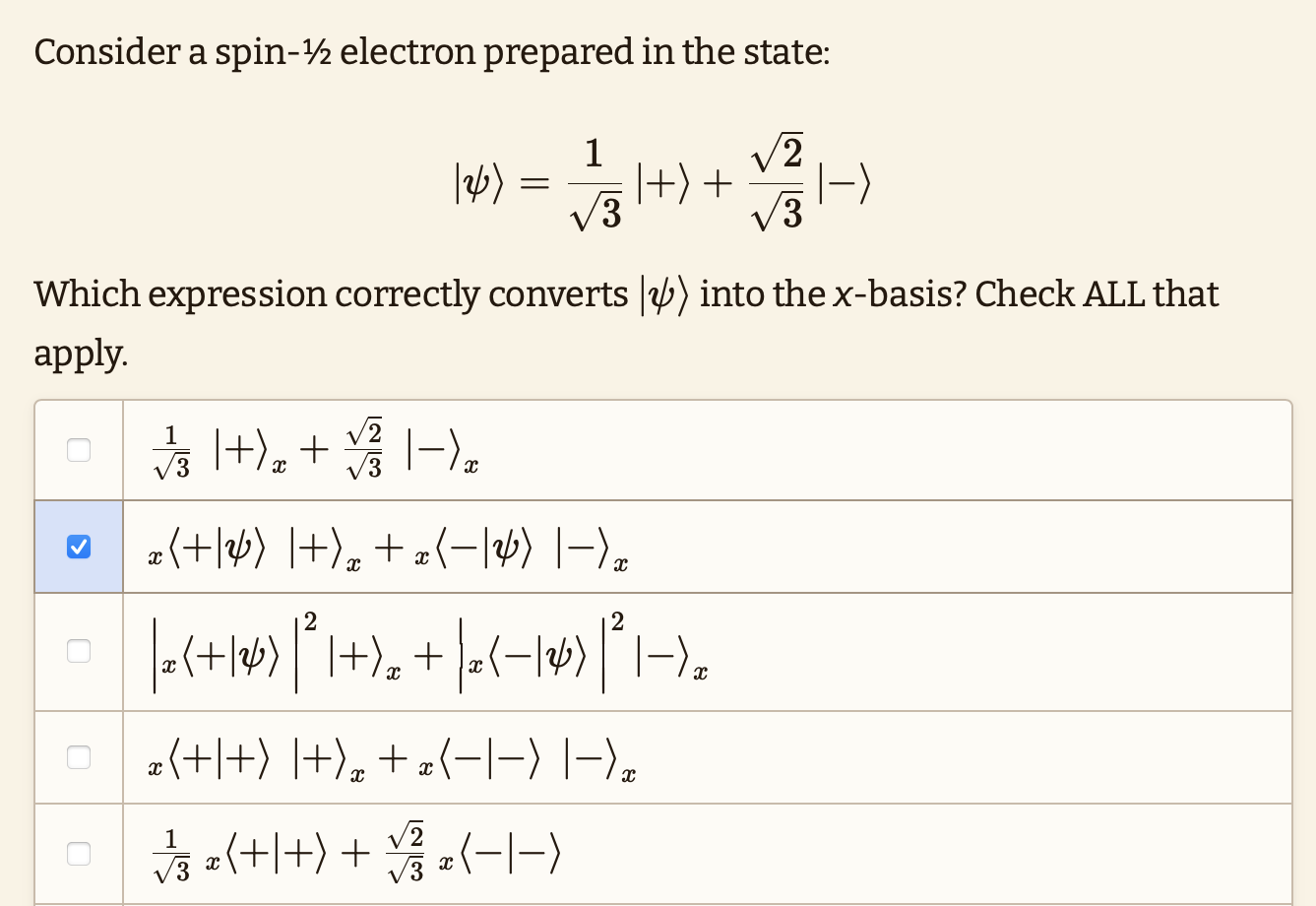}}%
  }%

    \caption{A screenshot of Pre-test Question 1 from the \censorace{ACE Physics} QBT.  The sole correct answer is checked.  The pre-test is included as the first page of the tutorial on the \censorace{ACE Physics} website.  The user interface in the rest of the tutorial looks similar to this screenshot.\label{fig:pre-test-q1}}

\end{figure}

In Fall 2021, the QBT was the only \censorace{ACE Physics} tutorial assigned.  To receive credit for the completing the tutorial, students provided answers to two tutorial questions that were repeated verbatim on the homework assignment, one of which was Pre-test Question 1 (\cref{fig:pre-test-q1}), described below.

In Spring 2022, the QBT was the third
\censorace{ACE Physics} tutorial assigned as homework. %---two 
%prior homework assignments had included different \censorace{ACE Physics} tutorials.
Additional \censorace{ACE Physics} tutorials were also assigned later in the semester.  A new feature in the platform provided the instructor with an estimate of how much of the tutorial each student had completed, which was used to assign credit for the homework problem.  Of all the students who accessed the QBT, only one did not reach the completion threshold required to receive full credit.

\textbf{Data collection:}
Data were collected in two settings: directly via the \censorace{ACE Physics} website, and on exams.  The \censorace{ACE Physics} QBT begins with a page entitled ``Before You Start'', which includes several pre-test questions.  Once a student moves on from the pre-test to the remainder of the tutorial, the \censorace{ACE Physics} system disallows changing pre-test answers.  The pre-test page instructs students to work alone, but we cannot confirm that students did not collaborate.  We might expect collaboration to inflate pre-test scores.

The pre-test includes two questions relevant to this study.  Pre-test Question 1 (\cref{fig:pre-test-q1}), which targets LG2, asks students to identify the expression that converts a ket written in the $z$-basis to the $x$-basis from a list of expressions including a variety of inner products.  Students were allowed to select multiple expressions, but a student's response was only considered correct if they exclusively selected the lone correct expression.

\begin{figure}[b]
  {%
  \setlength{\fboxsep}{0pt}%
  \fbox{\includegraphics[width=.99\linewidth]{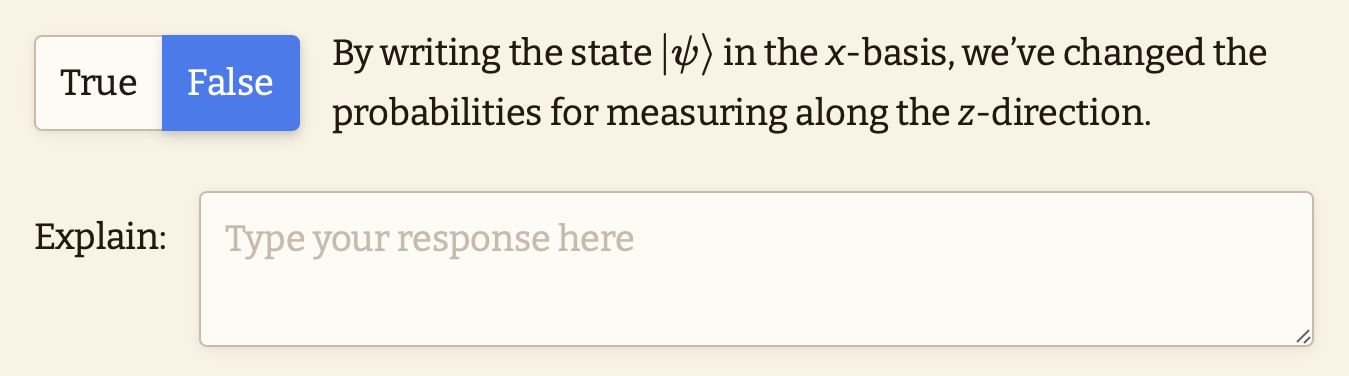}}%
  }%
  \caption{A screenshot of Pre-test Question 2 from the \censorace{ACE Physics} QBT.  The correct answer (False) is selected.  Students were only marked as correct if their response included a valid explanation.\label{fig:pre-test-q2}}
\end{figure}

Pre-test Question 2 (\cref{fig:pre-test-q2}), which targets LG1, asks students to consider whether changing basis affects measurement probabilities.  A student's response was considered correct if they selected ``False'' and also provided a correct explanation.  This question was repeated on the exams in Fall 2021 and Spring 2022, allowing for direct pre-/post-tutorial comparison.

One exam in each of the three semesters also included the exam question shown in \cref{fig:exam-q}.  The exam question, which targets LG2, asks students to convert a state expressed in the $z$-basis to the $n$-basis, defined by a pair of basis states provided in the question text.  Although students' exam grades included partial credit, for the purpose of this study we only considered a student's response correct if they provided completely correct coefficients for the state in the new basis.  We also coded students' exam responses according to the method they applied to change basis, specifically tracking whether they used projection inner products (indicated by expressions such as $a = {}_n{\braket{+|\psi}}$) versus any another method.  (These results were previously presented in~\censorcite{BasisPaper2} for the Fall 2018 exam data.)

Due to various curricular constraints, the exam question was given on the final exam in Fall 2018, on the first midterm in Fall 2021, and on the second midterm in Spring 2022.  Because of the continued relevance of basis change and projection inner products throughout the semester, we would not expect performance on this question to worsen as the semester progresses (we might expect performance to improve).  The Fall 2021 and Spring 2022 exams were held after the homework assignment that included the \censorace{ACE Physics} QBT was due.

\begin{figure}
  \begin{longfbox}[padding=2\fboxsep]
  \justifying
  Consider a spin-\nicefrac{1}{2} particle prepared in a the state $\ket{\psi}$, written in the $z$-basis as
  $\ket{\psi} = \frac{1}{\sqrt{2}} \ket{+} + i\frac{1}{\sqrt{2}} \ket{-}$.
  Write the state in the $n$-basis assuming
  \begin{equation*}
      \ket{+}_n = \frac{1}{\sqrt{3}} \ket{+} + i \frac{\sqrt{2}}{\sqrt{3}} \ket{-}
      \text{ and }
      \ket{-}_n = \frac{\sqrt{2}}{\sqrt{3}} \ket{+} - i \frac{1}{\sqrt{3}} \ket{-}
  \end{equation*}
  \textit{Hint: Solve for the values of $a$ and $b$ such that \\  $\ket{\psi} = a\ket{+}_n + b\ket{-}_n$}.
  \end{longfbox}
  \caption{One version of the Exam Question, which asked students to convert a spin-\nicefrac{1}{2} state to a different basis.  Versions of this question were asked on exams in three semesters.  Different semesters provided different coefficients for the $\ket{\pm}_n$ states, but all semesters included an imaginary coefficient.
  % (The state $\ket{\psi}$ is equal to $\ket{+}_y$, but this fact is not particularly useful when performing the calculation.)
  The exams in Fall 2021 and Spring 2022 also repeated Pre-test Question 2 (\cref{fig:pre-test-q2}) immediately after this change of basis question.
  Figure reproduced from~\censorcite{BasisPaper2}.\label{fig:exam-q}}
\end{figure}

% This study aims to evaluate the effectiveness of the \censorace{ACE Physics} QBT, so w
We will only report data for students who both completed the \censorace{ACE Physics} QBT and took the exam.  We have matched data for $N=66$ students from Fall 2021 and $N=92$ students from Spring 2022.  These account for the vast majority of students in both semesters.
In Fall 2018, $N=63$ students took the final exam, and none did the QBT.

\section{Results}

\cref{tab:exam-responses} reports student performance on the change of basis exam question from three similar semesters of upper-division quantum mechanics.  Comparison of results across semesters will allow us to evaluate the effect of the tutorial (with caveats discussed below).  We will compare (a) the number of students using projection; and (b) the number of students who answered the question correctly.  We will report statistical significance ($p$, computed with a chi-squared test) and effect size ($V$, computed via Cram\'er's V \cite{SciPyCramersV}).

\begin{table}[htbp]
  \caption{The methods students used to change basis on the exam question shown in \cref{fig:exam-q} from all three semesters.  In each row, the numbers in parentheses give the percentage of all students who both applied the given method and also produced the correct answer.  The Fall 2018 data were previously reported in~\censorcite{BasisPaper2}.\label{tab:exam-responses}}
  \begin{ruledtabular}
    \begin{tabular}{rccc}
                              & Without tutorial   & \multicolumn{2}{c}{With tutorial} \\
                              & Instructor A & Instructor A & Instructor B \\
                              & $N = 63$           & $N = 66$           & $N = 92$ \\
      \textbf{Applied method} & \textbf{Fall 2018} & \textbf{Fall 2021} & \textbf{Spring 2022} \\
      \hline
      Projection (correct/$N$) & 57\% (42\%)       & 71\% (61\%)        & 87\% (70\%) \\
      Other (correct/$N$)      & 43\% (8\%)        & 29\% (6\%)         & 13\% (3\%)
    \end{tabular}
  \end{ruledtabular}
\end{table}

A goal of the QBT is for students to apply the projection method when changing basis (LG2).  In Fall 2018, without the tutorial, 57\% of students used projection, whereas 71\% and 88\% of students used projection in Fall 2021 and Spring 2022, respectively (both later semesters included the QBT). The most direct comparison can be made between Fall 2018 and Fall 2021, because both were taught by the same instructor.  The increase in the fraction of students using projection from Fall 2018 to Fall 2021 is not statistically significant given this sample size ($p = 0.095$) and shows a weak effect size ($V = 0.15$).  The increase from Fall 2018 to Spring 2022 is significant ($p < 0.01$) with a moderate effect size ($V = 0.34$).  The difference in the fraction of students using projection between the two semesters with the \censorace{ACE Physics} QBT (Fall 2021 and Spring 2022) is also significant ($p < 0.05$) with a moderate effect size ($V = 0.20$), suggesting that factors beyond the QBT account for the higher rate of projection in Spring 2022.

Overall, 50\% of students answered the change of basis exam question correctly in Fall 2018, 67\% in Fall 2021, and 73\% in Spring 2022.  The improvement from Fall 2018 to the other two semesters is significant in both cases ($p < 0.05$), but we cannot necessarily attribute these improvements exclusively to the QBT.  The QBT focuses on the projection method for changing basis.  While the correctness rate for the projection method was higher in both Fall 2021 and Spring 2022 as compared with Fall 2018, the correctness rate for students using other methods was also marginally higher.  (The correctness rate is the percentage of students who applied a method correctly divided by the percentage who applied it at all.)

In addition to comparisons across semesters, we can also compare student performance on questions asked before and after they completed the \censorace{ACE Physics} QBT.  Pre-test Question~1 (\cref{fig:pre-test-q1}) was administered to students immediately before completing the tutorial and asked students to identify the expression with the correct inner products to convert a state from the $z$-basis to the $x$-basis.   Because identification of the correct inner products is a necessary step in applying the projection method to change basis, we consider these percentages an upper bound on the fraction of students who could successfully apply the projection method at the time of the pre-test.

We did not administer Pre-test Question 1 a second time in either Fall 2021 or Spring 2022; however, we consider the change of basis exam question as a post test.  Specifically, we can compare the number of students answering Pre-test Question 1 correctly against the number of students who used the projection method on the change of basis exam question and, at most, made a minor error.  Minor errors are limited to sign errors or simple arithmetic errors; all students who made at most a minor error used the correct inner product expressions.  It is possible that students using other methods on the exam might also have been able to apply projection, so these numbers are a lower bound on the fraction of students who could apply the projection method at the time of the exam.

\begin{table}[t!]
  \caption{Percentage of students answering Question 1 (\cref{fig:pre-test-q1}) and Question 2 (\cref{fig:pre-test-q2}) correctly on the pre-test (given immediately before the tutorial) and on the exam (given weeks after the tutorial) from the two semesters in which the tutorial was assigned.  Question 1 was not repeated directly on the exam; as the post-test percentage, we report the fraction of students that applied the projection method with at most minor errors on the change of basis exam question.\label{tab:pre-test-responses}}
  \begin{ruledtabular}
    \begin{tabular}{rccc}
         & \textbf{Pre-test}
         & \shortstack{\textbf{Direct}\\\textbf{post-test}} 
         & \shortstack{\textbf{Indirect}\\\textbf{post-test}}  \\
      \hline
      \multicolumn{3}{l}{\textbf{Fall 2021} ($N = 66$)} \\
      Question 1 & 46\%             & ---        & 68\% \\
      Question 2 & 65\%             & 97\%       & ---  \\
      \multicolumn{3}{l}{\textbf{Spring 2022} ($N = 92$)} \\
      Question 1 & 47\%             & ---        & 83\% \\
      Question 2 & 78\%             & 97\%       & --- 
    \end{tabular}
  \end{ruledtabular}
\end{table}

Before completing the \censorace{ACE Physics} QBT, 46\% and 47\% of students answered Pre-test Question 1 correctly in the Fall 2021 and Spring 2022 semesters, respectively (\cref{tab:pre-test-responses}).   After completing the QBT---and other relevant instruction---68\%  and 83\% of students applied the projection method correctly or with minor errors on the exam question in Fall 2021 and Spring 2022, respectively.  The improvements in both semesters are significant ($p < 0.01$) with moderate effect sizes ($V = 0.23$ in Fall 2021  and $V = 0.35$ in Spring 2022).

\cref{tab:pre-test-responses} also presents the percentages of students who answered Pre-test Questions 2 (\cref{fig:pre-test-q2}) correctly both on the tutorial pre-test and on the exam from both semesters.  The improvements between the pre-test and exam are significant in both semesters ($p < 0.01$) with moderate effect sizes ($V = 0.39$ in Fall 2021 and $V = 0.21$ in Spring 2022).  The pre-test scores were already fairly high---65\% in Fall 2021 and 78\% in Spring 2022.  Nonetheless, following the \censorace{ACE Physics} QBT---and other relevant instruction---nearly every student demonstrated an understanding in line with LG1.

\section{Discussions \& Conclusions}

We assigned the online \censorace{ACE Physics} QBT as homework in two semesters of upper-division quantum mechanics, and we administered pre- and post-tests targeting both learning goals.  We observed statistically significant improvements in student performance on the post-tests versus the pre-tests; however, we cannot isolate the effect of the QBT because other relevant instruction occurred between the pre- and post-tests.  We also compared student performance on an exam question targeting LG2 against exam data collected in a semester that did not include the QBT but was otherwise quite similar.  We observed a significant increase in the number of students using projection to change basis between the semester without the QBT and one of the semesters with the QBT---Spring 2022, which was taught by a different professor.  We did also observe an increase in the number of students using projection between the non-QBT semester and the other semester with the QBT---Fall 2021, which was taught by the same professor---but the difference was significant only at the $p < 0.1$ level.

What explains the discrepancy between the Fall 2021 and Spring 2022 semesters?  It is impossible to fully characterize the effects of different instructors, but one notable difference is that the Spring 2022 professor emphasized the derivation of the projection method for changing basis via the insertion of the identity operator $\ket{+}\!\bra{+}\,+\,\ket{-}\!\bra{-}$, whereas the Fall 2021 professor did not.  This form of the projection method appeared in over a quarter of Spring 2022 exam responses, whereas it only appeared twice in the Fall 2021 exam data.

How we administered the \censorace{ACE Physics} QBT also differed between semesters: the QBT was the only \censorace{ACE Physics} tutorial used in Fall 2021, whereas it was the third assigned in Spring 2022.  Moreover, a more robust mechanism was used to assign homework credit for tutorial completion in Spring 2022.  Students may have been better incentivized to engage with the \censorace{ACE Physics} QBT in Spring 2022, as well as better prepared to do so thanks to increased familiarity with the platform.  Student buy-in is relevant for any tutorial implementation, and our ongoing work investigates student engagement in the online, outside-class tutorial setting.

Overall, we consider our results sufficient to justify the assignment of the \censorace{ACE Physics} QBT as a single homework problem.  The cost-benefit analysis would differ if administering the QBT required an entire lecture period---indeed, given the relatively high pre-test scores, we have not elected to use the on-paper version of the QBT at our institution.  This is the exact goal of \censorace{ACE Physics}: to reduce the costs of administering tutorials while preserving some benefits.  Significant future research is required to assess the extent to which \censorace{ACE Physics} achieves this goal more broadly (including multiple tutorials and with different student populations), and we also continue to improve the platform.  Nonetheless---from the three courses considered in this study---the winning combination for teaching change of basis in a spins-first quantum mechanics class included assigning the \censorace{ACE Physics} QBT as homework and also emphasizing the insertion of the identity operator as a tool for understanding the projection method for basis change.

\acknowledgments{We thank \censor{Benjamin P. Schermerhorn, Homeyra Sadaghiani, and Gina Passante} for their contributions to the QBT and surrounding research.  This material is based upon work supported by the National Science Foundation under Grant \censor{No. PHY-2012147 as well as DUE 1626280, 1626594, and 1626482}.}

\bibliography{bibliography}

\end{document}